\documentstyle[preprint,aps,epsfig]{revtex}
\tightenlines

\begin{document}
\draft
\preprint{HUPD-9922}

\title{Dynamical fermion masses under the influence of \\
Kaluza-Klein fermions in extra dimensions}

\author{Hiroyuki Abe, Hironori Miguchi and Taizo Muta}
\address{Department of Physics, Hiroshima University \\
Higashi-Hiroshima, Hiroshima 739-8526, Japan}

\date{\today}

\maketitle

\begin{abstract}

  The dynamical fermion mass generation in the 4-dimensional brane 
  is discussed in a model with 5-dimensional Kaluza-Klein fermions 
  in interaction with 4-dimensional fermions. It is found that the 
  dynamical fermion masses are generated beyond the critical radius 
  of the compactified extra dimensional space and may be made small
  compared with masses of the Kaluza-Klein modes.
\end{abstract}

\narrowtext


\section{Introduction}	

\noindent

It is an interesting idea to assume an existence of the 
extra-dimensional space which eventually compactifies 
leaving our 4-dimensional space-time as a real world\cite{KK}. 
The recent proposal\cite{Ant,Ark1} for the mass scale of the compactified 
space to be much smaller than the Planck scale gave a strong 
impact on the onset of studying phenomenological evidences of 
extra-dimensional effects.
In recent approaches with extra dimensions it is usually assumed
that the standard model particles reside in the 4-dimensional brane
while the graviton may move around the bulk, the space-time with 
extra dimensions. In our present analysis we introduce bulk fermions 
in addition to the graviton and see what effects could be observed 
on the standard model particles. The bulk fermions interact with 
themselves as well as with fermions in the 4-dimensional brane 
through the exchange of the graviton and its Kaluza-Klein excited 
modes\cite{Han}, or through the exchange of gauge bosons which may be 
assumed to exist in the bulk\cite{Chen}. The interactions among 
fermions generated as a result of the exchange of all the Kaluza-Klein 
excited modes of the graviton or gauge bosons may be expressed 
as effective four-fermion interactions\cite{Han,Chen}. 
According to the four-fermion interactions we expect that 
the dynamical generation of fermion masses will take place.

In the present communication we look for a possibility of the dynamical 
fermion mass generation under the influence of the bulk fermions through 
the effective four-fermion interactions.
Although our argument is applicable to any higher dimensional models,
we confine ourselves to the 5-dimensional space-time for the convenience 
of explanations. In 5 dimensions fermion mass terms are forbidden if we 
require the symmetry under the chiral projection. The possible source of 
fermion masses in 4 dimensions is two-fold, i. e. the dynamically 
generated fermion masses and masses of the Kaluza-Klein excited modes
of the bulk fermions. The mass of the Kaluza-Klein excited modes is
known to be of order $1/R$ where $R$ is the radius of the compactified 
fifth dimension. 

We show in Sec. \ref{sec:MSi4DST} that the mixing between 
the brane fermion and 
bulk fermion does not lead to the large mass of order $1/R$ for the 
fermion in 4 dimensions.
We then consider in Secs. \ref{sec:EPfCF} and 
\ref{sec:GoDFM} whether the dynamically generated 
fermion mass can be made small compared with the mass of the Kaluza-Klein 
excited modes.
We calculate an effective potential for a composite operator composed
of a fermion and an anti-fermion in the leading order of the $1/N_f$ 
expansion with $N_f$ the number of fermion species. 
We find that the mass of fermions in the 4-dimensional brane is
generated dynamically if the compactification radius $R$ passes its 
critical value $R_C$ and the phase transition associated with the mass 
generation is of second order. This means that the fermion masses in the 
4-dimensional brane is small as far as the radius of the compactified 
fifth dimension is close to its critical value. 

\section{5-Dimensional Fermion Theory and Torus Compactification}

\noindent

We assume an existence of 5-dimensional bulk fermions $\psi$ 
in interaction with fermions $L$ on the 4-dimentional brane.
Effective interactions among these fermions can be given in the form 
of the four-fermion interaction. We imagine that such effective
interactions originate from the exchange of the bulk
gravitons between fermions. In fact it is known that the exchange 
of the Kaluza-Klein excited modes of the bulk graviton results in 
effective four-fermion interactions\cite{Han}. 
After the Fierz transformation on the four-fermion interactions 
we generate the transition-type interactions.
Accordingly we start with the following Lagrangian for our model 
\begin{equation}
   {\cal L\/}^{(5)}  
         =  \bar{\psi} i \gamma^M \partial_M \psi  \, + \, [ \,
             \bar{L} i \gamma^\mu \partial_\mu L 
              \, + \, g^2 \bar{\psi} \gamma^M L  \, 
                        \bar{L} \gamma_M \psi \, ] \, \delta( x^4 ),    
\label{eq:sLag}
\end{equation}
where $g$ is the coupling constant with mass dimension -3/2 and index $M$
runs from 0 to 4 while index $\mu$ runs from 0 to 3. 
Fermions $\psi$ and $L$ are assumed to be of $N_f$ components.

It is easy to see that the Lagrangian is symmetric under the chiral 
projection $x^4 \to -x^4$, $\psi(-x^4) \to -i\gamma^4 \psi(x^4)$ 
and $L(-x^4) \to -i\gamma^4 L(x^4)$. 
Thus, if we impose this symmetry on a Lagrangian describing our fermion 
system, any fermion mass terms are forbidden thus resulting in the above 
Lagrangian.
It should be noted here that, as is easily understood by referring to the 
Clifford algebra, an irreducible representation of the 5-dimensional 
fermion field is given by a 4-component field just as in the case of 
4 dimensions.
Hence we can use the same field both for the 4 and 5 dimensions. 
In odd dimensions it is well-known that there exists no object like 
$\gamma_5$ in 4 dimensions which commutes with all the $\gamma$ matrices.
Hence we do not have such object in 5 dimensions while the fifth component
of the $\gamma$ matrices in 5 dimensions, $\gamma^4$, turns out to be 
$i\gamma_5$ in 4 dimensions.

For the later convenience we rewrite the above Lagrangian by using auxiliary 
field $\sigma_M$ in the following equivalent form,
\begin{eqnarray} 
   {\cal L\/}^{(5)}  
          =  \bar{\psi} i \gamma^M \partial_M \psi  \,
              + \,[ \, 
              \bar{L} i /\!\!\!\partial L 
              \, - \, \sigma^M \, \sigma^*_M
              \, + \,  (g \, \sigma_M \bar{\psi} \gamma^M  L \, 
                 \, + \, h.c.)\, ]\, \delta(x^4). 
\end{eqnarray}
Since we are mainly interested in the dynamical fermion mass generation
in the leading order of the $1/N_f$ expansion, we neglect the irrelevant terms
in the Lagrangian by assuming $\left\langle \sigma_\mu \right\rangle=0$ where 
$\mu$ runs from 0 to 3.
After chiral rotation $\psi \to e^{i \frac{\pi}{4} \gamma_5} \psi$ and
$L \to e^{i \frac{\pi}{4} \gamma_5}L$ we have
\begin{eqnarray} 
   {\cal L\/}^{(5)}  
       = \bar{\psi} i /\!\!\!\partial \psi  \, - \,
              i \bar{\psi} \partial_4 \psi \, + \,[ \, 
              \bar{L} i /\!\!\!\partial L 
              \, - \, \left| \sigma \right|^2 
              \, + \,  (g \, \sigma  \, \bar{\psi}  L \,
                 \, + \, h.c.)\, ]\, \delta(x^4), \label{L}
\end{eqnarray}
where $\sigma=-\sigma_4$.
The Lagrangian (\ref{L}) is considered to be a composite-Higgs version of 
the mixing interaction adopted in Ref. \cite{Ark2,Die,Das}.

We now consider that the space of the fifth dimension is compactified 
on a circle with radius $R$. By adopting the periodic boundary condition
at $x_4=0$ and $x_4=2\pi R$ the bulk fermion field is expressed as a Fourier 
series,
\begin{equation}
  \psi(x^{\mu} ,x^{4}) = N \sum_{n=- \infty}^{\infty} 
                  \psi_n(x^{\mu}) \ e^{i \frac{n}{R} x^{4}},
\end{equation}
where $N$ is the normalization constant.

We define the 4-dimensional Lagrangian in the following way,
\begin{eqnarray}
  {\cal L\/}^{(4)} 
    &\equiv& \int_0^{2 \pi R} dx^4 {\cal L\/}^{(5)}    \nonumber \\ 
    &=& \sum_{n=-\infty}^{\infty}      
             \bar{\psi}_{n} i /\!\!\!\partial \psi_{n}      
        \ + \  \sum_{n=-\infty}^{\infty}      
                 \frac{n}{R} \bar{\psi}_{n} \psi_{n} 
        \ + \  \bar{L} i /\!\!\!\partial L \nonumber \\ &&
        \ - \  \left| \sigma \right|^2 
        \ + \ \left( \,m \sum_{n=-\infty}^{\infty}       
                   \, \bar{\psi}_{n} L     
        \ + \ h.c.\right),
\label{4Lag}
\end{eqnarray}
where $m \equiv N g \sigma$.
Note that we have normalized the kinetic terms in Eq. (\ref{4Lag}) by 
choosing $N \equiv 1 / \sqrt{2 \pi R} $.

\section{Mass Spectrum in 4-Dimensional Space-time}\label{sec:MSi4DST}

\noindent

In the following arguments we employ the matrix expressions
\begin{eqnarray}
 \Psi^t \equiv 
       \left( L \, , \, \psi_{0} \, , \, \psi_{1} \, , \,
           \psi_{-1} \, , \, \psi_{2} \, , \, \psi_{-2} \cdots
        \right) , \label{Psi}
\end{eqnarray} 
and
\begin{eqnarray}
  M \equiv      \left(
         \begin{array}{ccccccc}
            0 & m^{*} & m^{*} & m^{*} & m^{*} & m^{*} & \cdots \\
            m & 0 & 0 & 0 & 0 & 0 & \cdots \\ 
            m & 0 & \frac{1}{R} & 0 & 0 & 0 & \cdots   \\
            m & 0 & 0 & - \frac{1}{R} & 0 & 0 & \cdots   \\
            m & 0 & 0 & 0 & \frac{2}{R} & 0 & \cdots   \\
            m & 0 & 0 & 0 & 0 & - \frac{2}{R} & \cdots   \\          
            \vdots & \vdots & \vdots & \vdots & \vdots & \vdots & \ddots  
         \end{array}
        \right). \label{M}
\end{eqnarray}  
By using Eqs. (\ref{Psi}) and (\ref{M}) we rewrite the mass term and 
mixing term in the 4-dimensional Lagrangian such that
\begin{eqnarray}
  {\cal L\/}^{(4)}_{\rm mixing}    
      =  \sum_{n=-\infty}^{\infty}      
                 \frac{n}{R} \bar{\psi}_{n} \psi_{n} 
        \ + \ \left( \, m \sum_{n=-\infty}^{\infty}       
                   \,  \bar{\psi}_{n} L \, 
        \ + \ h.c. \right) \ 
      =\ \bar{\Psi} M \Psi.
\end{eqnarray}
If auxiliary field $\sigma$ acquires a non-vanishing vcuum expectation value,
we replace $\sigma$ in $m$ by its vacuum expectation value 
$\left\langle \sigma \right\rangle$, 
i. e. $m = N g \left\langle \sigma \right\rangle$. 
The eigenvalues of matrix $M$ determine the masses of 4-dimensional fermions. 
The eigenvalue equation is given by\cite{Die}
\begin{eqnarray}
\lefteqn{\det ( M - \lambda I )} \hspace{-0.1cm} \nonumber \\
 &=& \left[ \ \prod_{j=1}^{\infty} 
     \left( \lambda^2 - \left( \frac{j}{R} \right)^2 \right) \right] \,
     \left[ \lambda^2 - |m|^2 - 2 |m|^2 \lambda^2 \,
            \sum_{l=1}^{\infty}\, \frac{1}{\lambda^2 - 
            \left( l/R \right)^2}  \right] \label{Eigen} 
  = 0. 
\end{eqnarray}
It should be noted that the solutions $\lambda=j/R$ obtained by setting the first
factor in the middle of Eq. (\ref{Eigen}) to vanish is not the eigenvalues of 
Eq. (\ref{Eigen}) since they are canceled by the same factor in the denominator
in the second factor.
The real eigenvalues are obtained from the second factor in the middle of 
Eq. (\ref{Eigen}).
The summation in the third term of the second factor in Eq. (\ref{Eigen}) can
be performed as it is the Fourier series expansion of the cotangent function 
and thus the eigenvalue equation reduces to the simpler form\cite{Die},
\begin{eqnarray}
\lambda R  =  \pi |mR|^2 \cot(\pi \lambda R).\label{EigenS}
\end{eqnarray}

We would like to see whether we have a possibility of getting the light solution
in the eigenvalue equation (\ref{EigenS}). We confine ourselves to the case 
$|m| \ll 1/R$. By expanding the solution of Eq. (\ref{EigenS}) in powers of 
$|mR|$ we obtain\cite{Die} 
\begin{eqnarray}
\lambda_{\pm 0} R \ &=& \ \pm \ |mR| \, \left( 1-\frac{\pi^2}{6} |mR|^2 
                     + {\cal O}(|mR|^4)  \right), \label{Eigen0} \\
\lambda_{\pm n} R \ &=& \ \pm \ n \, \left( 1 + \frac{|mR|^2}{n^2} 
                     + {\cal O}(|mR|^4)  \right) \qquad (n \ne 0) \label{Eigenn}.
\end{eqnarray}
Obviously we find from Eqs. (\ref{Eigen0}) and (\ref{Eigenn}) that the lightest 
eigenvalue is given by
\begin{eqnarray}
\lambda_{\pm 0} = \pm \ |m| \quad {\textrm{for}} \quad |m| \ll 1/R. 
\label{Mass}
\end{eqnarray}
Thus we conclude that within our scheme there is a possibility of having the 
light fermion masses which is much smaller than the mass of the Kaluza-Klein
modes of the bulk fermion. It should be noted here that in Eq. (\ref{Mass}) 
we have two kinds of fermions with the positive and negative mass. 
These fermions are, however, indistinguishable since they have exactly the 
same properties. The next step that we have to proceed is to show that 
this light fermion mass is obtained as a result of the dynamical mass 
generation mechanism and can really be small.

\section{Effective Potential for Composite Fields}\label{sec:EPfCF}

\noindent

We are now interested in the actual value of the lightest fermion mass 
in the 4-dimensional brane as is given in Eq. (\ref{Mass}). Thus 
we have to study the dynamical mechanism of generating the non-vanishing 
vacuum expectation value of composite field $\sigma$ which is involved in the 
expression $m = N g \left\langle \sigma \right\rangle$. For this purpose 
we would like to calculate the effective potential for composite field $\sigma$.

Our 4-dimensional Lagrangian (\ref{4Lag}) is rewritten with the matrix
representation introduced in the last section as follows,
\begin{eqnarray}
  {\cal L\/}^{(4)} 
      =   \bar{\Psi} (M + I i /\!\!\!\partial) \Psi 
                  \ - \  \left| \sigma \right|^2 .
\end{eqnarray}
The generating functional $Z$ for our system is given by
\begin{eqnarray}
    Z = \int [ {\cal D\/} \bar{\Psi} ] 
                  [ {\cal D\/} \Psi ]
                   [{\cal D\/} \sigma ]
                    [{\cal D\/} \sigma^* ] ~
                      e^{ i \int d^4 x {\cal L\/}^{(4)} }. \label{4LagM}
\end{eqnarray}
By performing the path-integration for fermion field $\Psi$ in Eq. (\ref{4LagM})
we find in the leading order of the $1/N_f$ expansion,
\begin{eqnarray}
        Z &=& \int [{\cal D\/} \sigma ]
                    [{\cal D\/} \sigma^* ] ~
                      e^{ - i \int d^4 x V(\sigma) }, \\
V(\sigma) & \equiv & |\sigma|^2 - \int \frac{d^4 k}{i (2 \pi)^4} 
                             \ln \det (M + I /\!\!\!k),
\label{Veff}
\end{eqnarray}
where $V$ is the effective potential for the auxiliary 
field $\sigma$.
By performing the Wick rotation in the momentum integration in Eq. (\ref{Veff})
we rewrite the effective potential such that
\begin{eqnarray}
V(\sigma) = |\sigma|^2 - \frac{1}{8 \pi^2}
     \int_0^{\Lambda^2} d(k_E^2)\, k_E^2 \ln \det (M^2 + I k_E^2),
\end{eqnarray}
where $k_E$ stands for the Euclidean momentum and $\Lambda$ is the momentum cut-off.
Note that in addition to the divergence in the momentum integration the divergence
in the Kaluza-Klein mode sum shows up in general. In our 5-dimensional model this
divergence is not present while, if we start from the space-time dimensions 
higher than six, the divergence does exist and we have to rely on the regularization
method developed recently in Ref. \cite{Ban}. 
The same argument applies to the case where we derived Eq. (\ref{EigenS}).

\begin{figure}[htbp]
   \centerline{\epsfig{figure=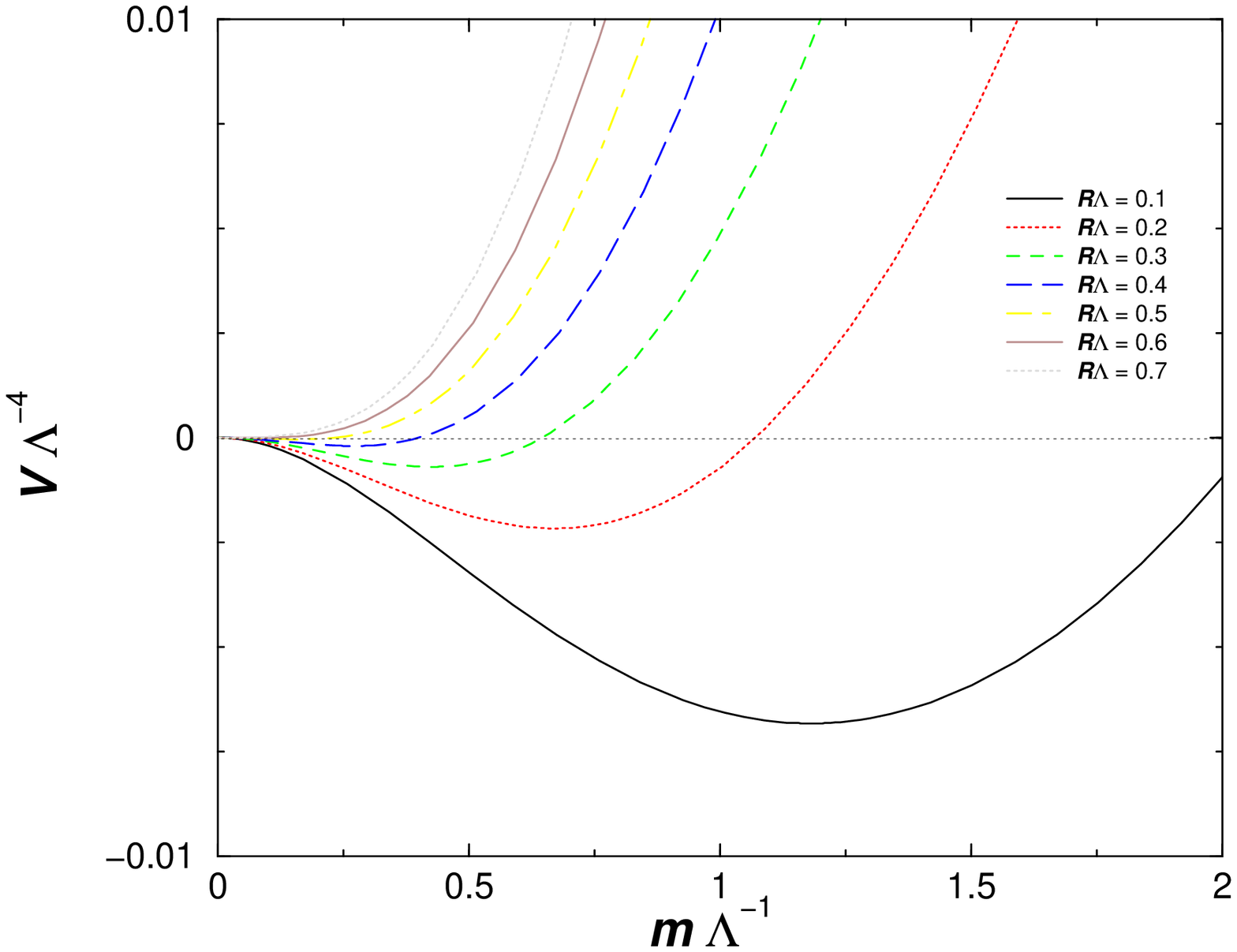,width=0.65\linewidth}}
   \caption{Typical behavior of the effective potential}
   \label{fig:Potential}
\end{figure}

After some calculations we find that
\begin{eqnarray}
V(\sigma)  =  |\sigma|^2 &-& \frac{1}{2 \pi^2}
                \int_0^\Lambda dx \, x^3
                \ln \left[x^2 + |m|^2 (\pi xR) \coth(\pi x R)\right] 
                \nonumber \\
          &-& \frac{1}{2 \pi^2} \sum_{j=1}^{\infty}\, 
                \int_0^\Lambda dx \, x^3
                \ln \left[x^2 + \left( \frac{j}{R} \right)^2\right].
\end{eqnarray}
The gap equation to determine the vacuum expectation value $<|\sigma|>$
of $|\sigma|$ reads
\begin{eqnarray}
\frac{\partial V(\sigma)}{\partial |\sigma|} 
   =  2 |\sigma| \, \left\{ 1- \frac{g^2}{2 \pi^2}\,
      \int_0^\Lambda dx \, \frac{x^3}{2 x \tanh(\pi x R) + g^2 |\sigma|^2}
      \right\} = 0.
\label{Gap}
\end{eqnarray}
By numerical observation of Eq. (\ref{Gap}) we find that there exists 
a non-trivial solution for $|\sigma|$ for a suitable range of parameters 
$g$ and $R$ and the solution corresponds to the true minimum of the 
effective potential. Accordingly the fermion mass is generated dynamically. 
Here the auxiliary field $\sigma$ (or the composite field $\bar{L}\psi$) 
acquires a vacuum expectation value.
The continuous symmetry that is broken by this dynamical process is the 
U(1) symmetry existed in the original Lagrangian (\ref{eq:sLag}): 
\begin{equation}
   \psi \to e^{i \beta} \psi, \ \ L \to e^{-i \beta} L.
\end{equation}
Moreover the phase transition associated with this symmetry breaking is of 
second order as is seen in Fig. \ref{fig:Potential}.

In the case of the second order phase transition the critical radius 
may be obtained by solving the equation derived from Eq. (\ref{Gap}) by 
setting $|\sigma|=0$,
\begin{equation}
 1- \frac{g^2}{4 \pi^2}\int_0^\Lambda dx \frac{x^2}{\tanh(\pi x R)} = 0.
\label{RC}
\end{equation}
Eq. (\ref{RC}) determines the relation between $R$ and $g$ with the 
suitable choice of the cut-off parameter $\Lambda$.

It is interesting to note here that, if we started from 3 dimensions 
instead of 5 dimensions and regarded the 2-dimensional world as a physical 
world, we would have obtained an equation for the critical radius corresponding 
to Eq. (\ref{RC}) as follows,
\begin{equation}
 1- \frac{g^2}{2 \pi}\int_\lambda^\Lambda dx \frac{1}{\tanh(\pi x R)} = 0,
\label{RC3}
\end{equation}
where parameter $\lambda$ is the cut-off of the low momentum integration
which is needed to remove an infrared divergence created by letting
$\sigma\to 0$.
The integration in Eq. (\ref{RC3}) is easily performed and we find that
\begin{equation}
\frac{g^2}{2\pi^2} = 
 R / \ln \frac{\sinh(\pi \Lambda R)}{\sinh(\pi \lambda R)}.
\label{RC3Sol}
\end{equation}
Eq. (\ref{RC3Sol}), if it is inverted, 
gives us a formula for the critical radius as a function of 
the four-fermion coupling constant. Similarly Eq. (\ref{RC}) gives us a relation
between $R$ and $g$ if we solve the equation numerically.

\section{Generation of Dynamical Fermion Masses}\label{sec:GoDFM}

\noindent

\begin{figure}[htbp]
   \centerline{\epsfig{figure=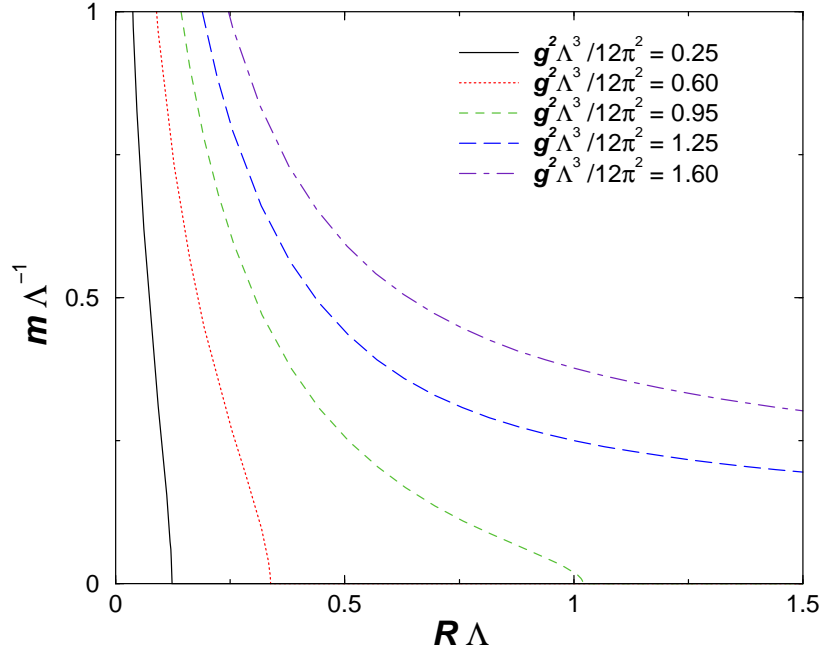,width=0.65\linewidth}}
   \caption{Dynamical fermion mass as a function of $R$ with $g$ fixed.}
   \label{fig:m-R_g}
\end{figure}

\begin{figure}[htbp]
   \centerline{\epsfig{figure=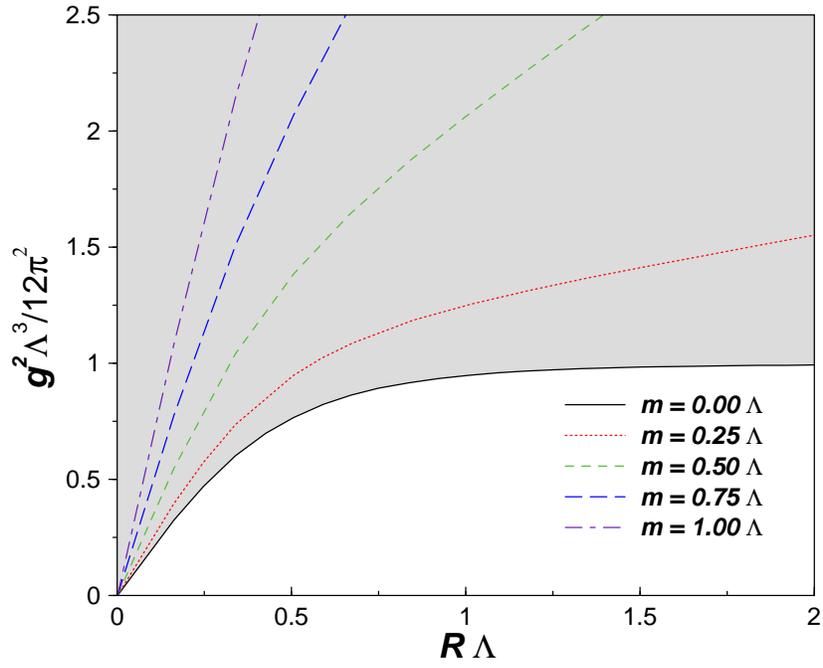,width=0.65\linewidth}}
   \caption{Critical radius as a function of $g$.}
   \label{fig:g-R}
\end{figure}

As was shown in Eq. (\ref{Mass}) the lowest fermion mass on the 
4-dimensional brane is $m=Ng <|\sigma|>$ where $<|\sigma|>$ is 
determined by solving Eq. (\ref{Gap}). By using the numerical estimation 
we calculate the fermion mass $m$ as a function of radius $R$ for fixed
coupling constant $g$. The result is shown in Fig. \ref{fig:m-R_g} where 
$m$, $R$ and $g$ are normalized by the cut-off parameter $\Lambda$.
It is immediately recognized that the fermion mass generation takes 
place below the critical radius and so is kept small near the critical radius.
The critical curve which represents the critical radius as a function of the 
coupling constant is shown in Fig. \ref{fig:g-R} as the curve for $m=0$. 
It is easy to see that $g^2 \Lambda^3 = 12 \pi^2$ as $R \to \infty $ on 
the critical curve.

In our model the bulk fermions reside in the 5-dimensional space-time
and their Kaluza-Klein modes show up as fermions with masses $n/R$ in the
4-dimensional space-time. Since those fermions are not observed in the 
present experimental situation except for the zero mode which mixes with 
the fermion on the brane, their masses have to be very high and hence 
$1/R$ should be much higher than several TeV. Thus the compactification 
scale $R$ for the fifth dimension is considered to be very small. On 
the other hand the bulk graviton lives in the higher dimensional space-time 
and the compactification scale $1/R_G$ for the extra-dimensional space 
associated with the bulk graviton could be smaller than $1/R$ as is
suggested in Ref. 2. Our cut-off scale $\Lambda$ is introduced 
to suppress the divergence appearing in the integration in the effective 
potential. Hence it is to be determined by the region of the validity of 
our effective four-fermion theory. We suppose that $\Lambda > 1/R$ since 
the fundamental theory which derives our effective theory is realized 
at much higher scale than the compactification scale for the bulk fermions.
In Fig. \ref{fig:g-R} we recognize that, if $g^2\Lambda^3 \sim 12\pi^2$, 
the fermion mass is kept small for a wide range of radius $R$ except for 
the small $R$ region.

\section{Conclusions}

\noindent

We have found within our model that in spite of the presence of the large 
mass scale $1/R$ in the theory the fermion masses on the 4-dimensional 
brane can be made small as a consequence of the interaction among 
the bulk and brane fermions: the mixing of the brane fermions with 
the bulk fermions does not lead to the lightest fermion masses 
of order $1/R$ and also the dynamically generated fermion masses 
are not of order $1/R$.
This result is obtained because the dynamical fermion masses generated 
under the second-order phase transition are small irrespectively of $1/R$ 
near the critical radius. In our model the possibility of having low mass 
fermions resulted from the dynamical origin. This mechanism is quite 
different from the ones in other approaches in which low mass fermions 
are expected to show up as a result of the kinematical 
origins\cite{Ark2,Die,Das,Moh,Yos}.

It is tempting to make a final comment on the conjecture that the nature 
chooses the compactification scale near the critical radius so as to 
have low mass fermions. In this connection it may be interesting to 
suppose that the existence of the critical radius is a result of the 
physical process that the compactification of the fifth dimensional space 
is driven by the force unknown to us and this force is balanced by the 
pressure coming from the fermionic Casimir energy\cite{Gun,Abe}.

\acknowledgments
The authors would like to thank Masahiro Yamaguchi (Tohoku U.) and 
Koichi Yoshioka (Kyoto U.) for fruitful discussions and correspondences.
They are also indebted to Tak Morozumi for useful comments and to
Keith Dienes, Emilian Dudas and Tony Gherghetta 
for a useful correspondence. The present work is supported financially by 
the Monbusho Grant, Grant-in-Aid for Scientific Research (C) with contract 
number 11640280.


\pagebreak

%


\end{document}